\documentclass[10pt,twocolumn]{revtex4}
\usepackage{epsfig}

\begin{document}

\title{Tunability of wire-grid metamaterial immersed into nematic liquid crystal}

\author{ M.V. Gorkunov$^{1,2}$ and M.A. Osipov$^1$}
\affiliation{$^1$Department of Mathematics, University of
Strathclyde, Glasgow G1 1XH, UK \\ $^2$Institute of Crystallography,
Russian Academy of Sciences, 119333 Moscow, Russia}

\begin{abstract}
We propose electrically tunable hybrid metamaterial consisting of
special wire grid immersed into nematic liquid crystal. The
plasma-like permittivity of the structure can be substantially
varied due to switching of the liquid crystal alignment by external
voltages applied to the wires. Depending on the scale of the
structure, the effect is available for both microwave and optical
frequency ranges.
\end{abstract}
\maketitle

Electromagnetic properties of periodic arrays of metal wires,
stripes and finely structured particles have attracted increasing
attention in recent years. The variety of emerging phenomena extends
from the negative refraction and subwavelength light constraining
\cite{Pendry2000} to extraordinary transmission \cite{Ebbesen},
which all have been attributed to regular metallic structures with
periods smaller or much smaller than the wavelength of
electromagnetic radiation. As a rule, the unusual properties are
characterized by a considerable or even anomalously strong frequency
dispersion. Undoubtedly, this opens prospective possibilities for
tuning and switching, since a small modification of the system may
result in a substantial change of its electromagnetic response at a
given signal frequency.

Liquid crystals (LC) inherently have all the necessary
characteristics to provide tunability to subwavelength metal arrays.
By switching the orientation of LC medium between metal elements or
in metal cavities one can trigger substantial changes in the overall
properties. Importantly, LCs are transparent and significantly
anisotropic in both microwave and optical frequency ranges, in which
subwavelength metal arrays are usually operating \cite{deJeu,
Sambles}. Finally, the metal constituents can be readily used as
electrodes for applying switching voltages.

In the past decade, various types of metamaterials have received
significant attention  due to negative values of permeability and
permittivity (see e.g. the recent reviews \cite{shalaev, linden} and
refs. therein). Certain methods are known how to provide tunability
to metamaterials at microwave frequencies by inserting electronic
semiconductor components \cite{lapin, zharov, padilla}. However, it
still remains challenging to achieve this at optical frequencies. In
this Letter we describe how nematic LC environment can provide
tunability to the plasma frequency of wire-grid metamaterial.

\begin{figure}
\centering \epsfig{file=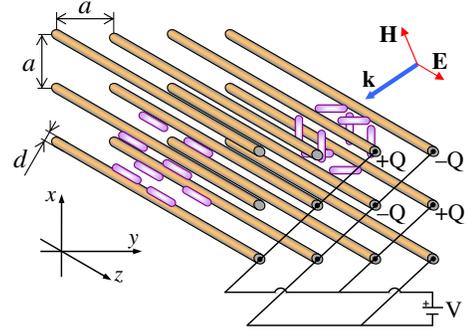, width=0.7\columnwidth}
\caption{(Color online) Schematic of switching of nematic LC
surrounding grid of wires. Note: the size of LC molecules is
extremely exaggerated.} \label{scheme}
\end{figure}

Wire lattices exhibit plasma-like permittivity for electromagnetic
radiation with wavelengths much larger than the lattice period
~\cite{Pendry,Belov}. In particular, for electromagnetic waves
traveling normally to the grid of infinite parallel wires and
polarized along them ($xy$-plane of incidence and $z$-polarization
of electric field in Fig.~\ref{scheme}), the employed component of
the effective local permittivity is
\begin{equation}\label{eps}
\varepsilon_{zz}(\omega)=\varepsilon^{(h)}_{zz}(\omega)-\frac{\Omega^2}{\omega^2}.
\end{equation}
Here ${\varepsilon}^{(h)}_{zz}$ is the $zz$-component of the
permittivity tensor of host medium surrounding the wires, and the
parameter $\Omega$ is determined by the grid geometry.

The so-called plasma-frequency, at which the permittivity
(\ref{eps}) changes its sign, is $\omega_0=\Omega/
(\varepsilon^{(h)}_{zz})^{1/2}$. The grid on its own is transparent
at frequencies higher than $\omega_0$ and reflecting at lower
frequencies. Being accompanied by a 3D array of split-ring
resonators providing negative permeability, the structure is a
transparent left handed medium for $\omega<\omega_0$ and reflects
waves with $\omega>\omega_0$.

The presence of ${\varepsilon}^{(h)}_{zz}$ in Eq.~(\ref{eps})
suggests an easy way of tuning the metamaterial by varying the host
medium permittivity. Nematic LCs provide a natural opportunity to
realize this. Anisotropic LC permittivity axes follow the
orientation of the director $\bf n$:
\begin{equation}\label{epsLC}
\varepsilon^{(h)}_{ij}(\omega)=\varepsilon_\bot(\omega)\
\delta_{ij}+\varepsilon_a(\omega)\  n_i n_j.
\end{equation}

Typically,  elongated molecules of nematic LC tend to align along
interfaces due to surface anchoring. The resulting LC alignment
along the wires is shown in Fig.\ref{scheme} on the left.
If external voltages are applied to the neighboring wires, this will
load them with charges of different signs. The arising static
electric field in the $xy$-plane will force the LC molecules to
orient perpendicularly to the wires as shown in Fig.\ref{scheme} on
the right. As a result, $\varepsilon^{(h)}_{zz}$  switches from
$(\varepsilon_\bot+\varepsilon_a)$ to $\varepsilon_\bot$ providing
the relative switching range of $\omega_0$
\begin{equation}\label{delom}
\frac{\Delta\omega_0}{\omega_0}\simeq\frac{\varepsilon_a}{2\varepsilon_\bot}.
\end{equation}
The dielectric anisotropy of LC is known to be at least of the order
of several dozens of percent for both microwaves and light
\cite{deJeu, Sambles}. Therefore, the proposed design provides a
possibility of shifting the plasma-frequency by 10--20\%.

Let us find out the voltage to be applied to the grid for efficient
switching. Consider the quadratic lattice of cylindrical wires as
shown in Fig.~\ref{scheme}. Assume that a voltage $V$ applied across
the nearest neighbors loads the wires with charge per length
densities $\pm Q$. The charges are distributed over the surfaces of
thin wires. We neglect the angular inhomogeneity of this
distribution and assume the electric field of a wire to be the same
as that of a charged wire axis. We also presume that the static
dielectric anisotropy of the nematic $\varepsilon_a^{\rm
st}=\varepsilon_a(0)$ is small and the electric field pattern is not
perturbed by the inhomogeneously oriented LC.

A single wire stretched along the line $x=0, y=0$ and loaded with
the charge $Q$ per unit length produces the electric field potential
$\varphi_1({ \rho})=-\frac{Q}{2 \pi \varepsilon_0 \varepsilon_{\rm
st}}\log (2\rho/d)$, where $\rho=\sqrt{x^2+y^2}$, $\varepsilon_{\rm
st}=\varepsilon_\bot(0)$ is the static permittivity of the LC, and
the zero of the potential is assigned to the wire surface. In the
infinite 2D lattice of alternatingly charged wires, the total
potential outside of the wires is
\begin{equation}
\varphi({\bf
r})=\sum_{n,m=-\infty}^{n,m=\infty}{(-1)^{n+m}\varphi_1\left(\rho_{nm}\right)},\label{phi}
\end{equation}
where $\rho_{nm}=\sqrt{(x-na)^2+(y-ma)^2}$ and the coordinate origin
lies on the axis of a positively charged wire.

We have calculated (\ref{phi}) for finite latices. In spite of the
absence of true mathematical convergence of the series, the
potential distribution establishes already for a lattice of
$10\times10$ wires. Further increase of the number of wires does not
affect the potential in the middle provided that the overall
electric neutrality is preserved.
\begin{figure}
\centering \epsfig{file=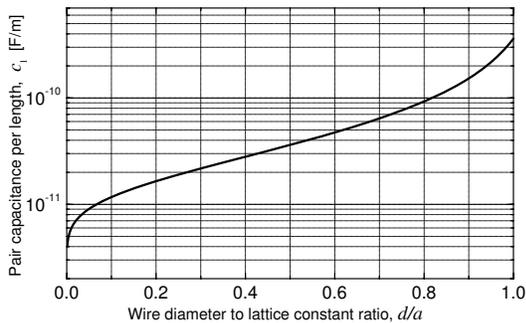, width=0.8\columnwidth}
\caption{Grid capacitance per pair of wires in vacuum per length
versus the ratio of wire diameter to lattice constant.}
\label{figC1}
\end{figure}

Applied voltage equals the difference of the potential on positively
and negatively charged wires, $V=\varphi_+-\varphi_-$. The
capacitance per length of a grid of large even number $N$ of wires
is $C=Q N/(2 V)$. It is convenient to introduce the capacitance per
pair of wires in the vacuum per length, $c_1=2C/(N\varepsilon_{\rm
st})$, which naturally appears in the relation $Q=\varepsilon_{\rm
st} c_1 V$. Apparently, $c_1$ is a function of two geometric
parameters, $a$ and $d$, and depends on their ratio $d/a$. The
calculated values of $c_1$ are presented in Fig.~\ref{figC1}. Note
that $c_1$ tends to infinity at $d\simeq 1.08 a$ instead of $d=a$,
when the neighboring wire surfaces come in contact. This is the
consequence of neglecting the anisotropy of the charge distribution
over wire surface. We see that even for thick wires our
approximation is reasonably good.

\begin{figure}
\centering \epsfig{file=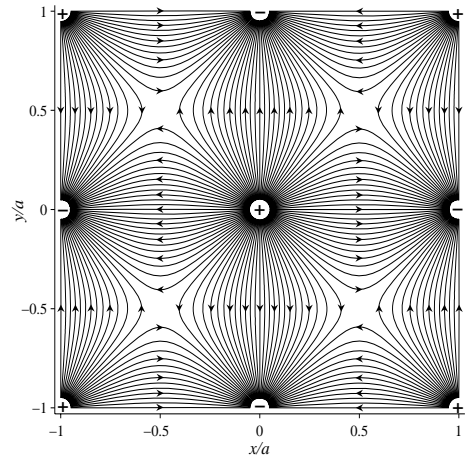, width=0.7\columnwidth}
\caption{Lines of electric filed produced by 2D grid of
alternatingly charged wires. The wire diameter is set to be 0.1 of
the lattice period} \label{figE}
\end{figure}

Typical calculated electric field pattern is shown in
Fig.~\ref{figE}. The field amplitude is maximum at the wire surfaces
(where the lines are condensed) and vanishes at the middle points of
the diagonal lines connecting the nearest wires with the same sign
of charge.

Now we consider the effect of this field on the nematic. If we
totally neglect the LC elasticity, the anisotropic molecules should
align along the field lines. The director $\bf n$ then lies strictly
in the $xy$-plane and the switching is perfect. The finite
elasticity of the LC disturbs the ideal picture and the bending
pattern of the director differs from the field lines pattern.
However, this is of minor importance for the
$\varepsilon^{(h)}_{zz}$ switching since the director still stays
within the $xy$-plane.

The critical effect of LC elasticity occurs in the vicinity of the
wires. To ensure the LC orientation along the wires when the voltage
is switched off, the anchoring of LC molecules at wire surfaces must
be strong. When the voltage is switched on, the anchoring still
forces the molecules at the surfaces to point along the $z$-axis.
Accordingly, the director has to rotate by an angle of $\pi/2$
within the transient layer from the wire surface to the LC bulk.
Efficient switching occurs when this layer is thin compared to the
scale of the structure (lattice constant), i.e., practically
important is the limit of strong electric field and thin transient
layer.

In this limit, we can neglect the contributions from other wires to
the field near a wire surface. In the cylindrical coordinates, the
only present $\rho$-component of the electric field equals
$E_\rho(\rho)=\frac{Q}{2 \pi \varepsilon_0 \varepsilon_{\rm
st}}\rho^{-1}$ while the nematic director has two components
$n_\rho(\rho)$ and $n_z(\rho)$. At the wire surface $n_\rho(d/2)=0$.

The free energy of the LC can be presented as the sum  $F=F_K+F_E$
of the elastic deformation energy and the dielectric energy
\cite{deJeu}, where the latter reads
\begin{equation}
F_E=-\pi \varepsilon_0 \varepsilon_a^{\rm st} \int_d^R \rho\ d\rho\
(E_\rho n_\rho)^2.
\end{equation}
Here the upper limit $R$ is large enough to assure $n_z(R)=0$ and
can be extended to infinity. The simplest form of the LC elastic
energy is given by the so-called one constant approximation:
\begin{equation}
F_K=\frac{K}{2}\int dV\ \left[(\nabla\cdot\bf n)^2+(\nabla\times\bf
n)^2\right].
\end{equation}

\begin{figure}
\centering \epsfig{file=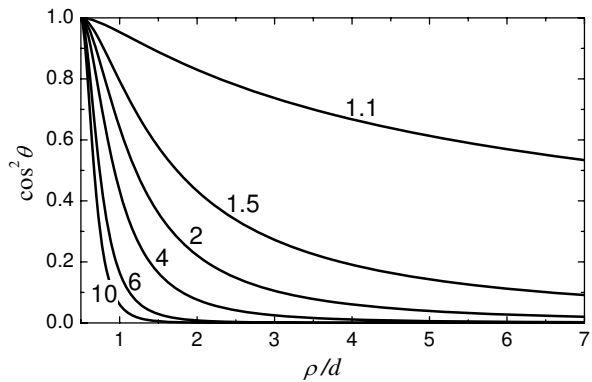, width=0.9\columnwidth}
\caption{Spatial profiles of $\cos^2\theta$ determining modulation
of LC permittivity $\varepsilon_{zz}$. Corresponding values of $v^2$
are shown.} \label{figtheta}
\end{figure}

Introducing the director polar angle $\theta$ as
$n_\rho=\sin\theta$, $n_z=\cos\theta$ we obtain the compact form of
the total energy
\begin{equation}\label{F}
F=\pi K \int_d^\infty \rho\ d\rho\
\left[(\theta')^2+(1-v^2)\frac{\sin^2\theta}{\rho^2}\right],
\end{equation}
where the parameter $v^2=V^2c_1^2\varepsilon_a^{\rm
st}/(4\pi^2K\varepsilon_0)$ characterizes relative contribution of
the voltage-driven term.

Minimizing the functional (\ref{F}) yields the differential equation
for $\theta(\rho)$:
\begin{equation}
\theta''+\frac{1}{\rho}\
\theta'+(v^2-1)\frac{\sin2\theta}{2\rho^2}=0
\end{equation}
with the boundary conditions $\theta(d/2)=0$ and
$\theta(\infty)=\pi/2$. For $v^2>1$ the exact solution reads:
\begin{equation}
\theta(\rho)=\frac{\pi}{2}-2 \arctan
\left[\left(d/2\rho\right)^{\sqrt{v^2-1}}\ \right].
\end{equation}

The spatial variation of the $zz$-component of dielectric tensor
(\ref{epsLC}), is controlled by the factor
$n_z^2(\rho)=\cos^2\theta(\rho)$. The profiles of the latter are
given in Fig.~\ref{figtheta} for several values of $v^2$ larger than
unity. It is seen that the voltage has to exceed considerably the
critical value to provide the LC reorientation within thin transient
layer. For $v^2\gtrsim2$ the layer thickness becomes comparable with
the wire diameter. For estimates, we take a nematic LC elastic
modulus $K\simeq10^{-11}$N, static permittivity $\varepsilon_{\rm
st}\simeq10$ and static dielectric anisotropy $\varepsilon_a^{\rm
st}\simeq1$. We also set the wire diameter to be ten times smaller
than the lattice constant, which according to Fig.~\ref{figC1}
yields $c_1\simeq10^{-11}$F/m. Then appropriate voltage is estimated
as
\begin{equation}
V=\frac{2\pi}{c_1}\sqrt{\frac{2K\varepsilon_0}{\varepsilon_a^{\rm
st}}}\simeq8.4 {\rm V}.
\end{equation}
Remarkably, this moderate voltage is independent of the scale of the
grid.

One notes that the studied LC switching geometry differs
qualitatively from the conventional one within flat electrooptic
cells. In our case, the LC bulk is directly aligned and reoriented
by the immersed wires. We believe this to be an additional advantage
for microwave applications, where the controlling of millimeter (or
even centimeter) thick LC samples by cell surfaces is practically
impossible.

In summary, we have demonstrated that the plasma frequency of
wire-grid metamaterial immersed into nematic LC can be efficiently
tuned by 10-20\% due to the switching of LC alignment. Our estimates
show that the necessary voltage applied to the wires is of the order
of several Volts. The main conclusions are valid for both microwave
and optical frequency ranges.


\begin{thebibliography}{10}



\bibitem{Pendry2000}
J.B.~Pendry, Phys. Rev. Lett. {\bf 85}, 3966 (2000).

\bibitem{Ebbesen}
W.L. Barnes, A. Dereux, and T.W. Ebbesen, Nature {\bf 424}, 824
(2003).

\bibitem{deJeu}
W.H. de Jeu, {\it Physical Properties of Liquid Crystalline
Materials}, Gordon and Breach, London, 1980.

\bibitem{Sambles}
J.R. Sambles, R.Kelly, and F. Yang, Phil. Trans. R. Soc. A {\bf
364}, 2733 (2006).

\bibitem{shalaev}
V.M. Shalaev, Nature Photonics, {\bf 1}, 41 (2007).

\bibitem{linden}
S. Linden, C. Enkrich, G. Dolling {\it et al.}, IEEE J. of Selected
Topics in Quant,. Electr., {\bf 12},  1097 (2006).

\bibitem{lapin} M. Lapine, M. Gorkunov, and K.H. Ringhofer, Phys. Rev. E {\bf 67}, 065601(R) (2003).

\bibitem{zharov} A.A. Zharov, I.V. Shadrivov, and Yu.S. Kivshar, Phys. Rev. Lett.
{\bf 91}, 037401 (2003).

\bibitem{padilla} H.-T. Chen, W.J. Padilla, J.M.O. Zide, {\it et al.}, Nature {\bf 444}, 597 (2006).

\bibitem{Pendry}
J.B. Pendry, A.J. Holden, W.J. Stewart, and I. Youngs, Phys.Rev.
Lett. {\bf 76}, 4773 (1996).

\bibitem{Belov}
P.A. Belov, R. Marques, S. I. Maslovski {\it et al.}, Phys. Rev. B
{\bf 67} 113103 (2003).


\end{thebibliography}
\end{document}